\documentclass[twocolumn,aps,pra,showpacs,superscriptaddress]{revtex4-1}

\usepackage{amssymb}
\usepackage{mathrsfs}
\usepackage{lipsum}
\usepackage{graphicx}
\usepackage[caption=false]{subfig}
\usepackage{mathtools}
\usepackage{epstopdf}
\usepackage{xcolor}
\usepackage[colorlinks, linkcolor=red, anchorcolor=green, citecolor=blue, urlcolor=violet]{hyperref}
\DeclareGraphicsExtensions{.pdf,.jpg,.png,.eps}
\usepackage[toc,page,title,titletoc,header]{appendix}
\begin{document}

\title{Toward demonstrating super-sensitive angular displacement estimation by using orbital angular momentum coherent state and SU(1,1)-SU(2) hybrid interferometer}

\author{Jian-Dong Zhang}
\affiliation{Department of Physics, Harbin Institute of Technology, Harbin, 150001, China}
\author{Zi-Jing Zhang}
\email[]{zhangzijing@hit.edu.cn}
\affiliation{Department of Physics, Harbin Institute of Technology, Harbin, 150001, China}
\author{Long-Zhu Cen}
\affiliation{Department of Physics, Harbin Institute of Technology, Harbin, 150001, China}
\author{Jun-Yan Hu}
\affiliation{Department of Physics, Harbin Institute of Technology, Harbin, 150001, China}
\author{Yuan Zhao}
\email[]{zhaoyuan@hit.edu.cn}
\affiliation{Department of Physics, Harbin Institute of Technology, Harbin, 150001, China}

\date{\today}

\begin{abstract}

In this paper, we propose a protocol for angular displacement estimation based upon orbital angular momentum coherent state and a SU(1,1)-SU(2) hybrid interferometer. 
This interferometer consists of an optical parametric amplifier, a beam splitter and reflection mirrors, hereon we use a quantum detection strategy $\---$ balanced homodyne detection. 
The results indicate that super-resolution and super-sensitivity can be realized with ideal condition. 
Additionally, we study the impact of photon loss on the resolution and the sensitivity, and the robustness of our protocol is also discussed.
Finally, we demonstrate the advantage of our protocol over SU(1,1) and summarize the merits of orbital angular momentum-enhanced protocol.

\end{abstract}

\pacs{42.50.Dv, 42.50.Ex, 03.67.-a}

%42.50.Dv Quantum state engineering and measurements 
%42.50.Ex Optical implementations of quantum information processing 
%and transfer 
%03.67.-a Quantum information  

\maketitle

\section{Introduction}
In recent years, apart from phase estimation, there has been a surge in interest in estimation of angular displacement as another degree of freedom and has received wide attention \cite{zhang2016ultra,PhysRevLett.81.4828, zhang2016super, cen2017state,PhysRevLett.112.200401, zhang2017effects, PhysRevA.83.053829,zhang2017improved,liu2017enhancement,zhang2018oam}. 
Angular displacement can be divided into two distinct physical scenarios, one is the rotation of the polarization \cite{cen2017state, zhang2017effects} and the other is the relative rotation of the light axis \cite{zhang2016ultra, PhysRevLett.81.4828,zhang2016super,zhang2017improved,liu2017enhancement,PhysRevLett.112.200401,PhysRevA.83.053829,zhang2018oam}. 
Rotation of the polarization is often caused by birefringence effect of wave plates or Faraday magneto-optical rotation effect of optical crystals. 
Rotation of the light axis can be introduced by the rotation of the prism or, equivalently, the rotation of the reference coordinate system \cite{d2013photonic}.

For the first case, Malus law and shot-noise limit (SNL) are the representations of the classical metrology. 
A great deal of work has been done to improve such sensitivity \cite{cen2017state, zhang2017effects, PhysRevLett.112.153601, Wolfgramm}. 
It has been shown that the sensitivity can be boosted by using N00N state  \cite{cen2017state} or two-mode squeezed vacuum \cite{zhang2017effects} along with quantum detection strategy. 
As to the latter one, also the angular displacement type we study in this paper, spin angular momentum (SAM) \cite{PhysRev.50.115} and orbital angular momentum (OAM) \cite{PhysRevA.45.8185} of light are usually utilized. 
The OAM performs better than SAM since SAM forms a two-dimensional Hilbert space (spin up or spin down), whereas, OAM space is inherently infinite dimensional.

Recently, Heisenberg-limited estimation protocols towards angular displacement based upon OAM-enhanced have been realized in both SU (1,1) and SU (2) interferometers \cite{zhang2017improved, liu2017enhancement}. 
Due to the difficulty in preparing exotic quantum state with large photon number, the estimation sensitivities using these quantum states are even inferior to that of high-intensity coherent state \cite{Escher2011, PhysRevLett.111.173601}.
Therefore, angular displacement estimations based upon coherent state in SU(2) interferometer have also been extensively studied, with the sensitivity being limited by SNL. 

On the other hand, some studies have shown that the sensitivity of the coherent state in SU (1,1) interferometer can surpass the SNL \cite{PhysRevA.85.023815, PhysRevA.86.023844}.
In this paper, we propose an OAM-enhanced protocol that utilizes OAM coherent state and a SU(1,1)-SU(2) hybrid interferometer to achieve sub-shot-noise-limited sensitivity, which is better than both SU(1,1) and SU(2) cases.

The remainder of this paper is organized as follows: 
in Sec. \ref{II}, we briefly introduce the structure and the working principle of the hybrid interferometer. 
Then, we discuss balanced homodyne detection with ideal situation in Sec. \ref{III}. 
In Sec. \ref{IV}, we study the effects of photon loss on resolution and sensitivity, additionally, we compare our protocol and the SU (1,1) one. 
Finally, we conclude our work with a summary in Sec. \ref{V}.

\begin{figure*}[htbp]
\centering
\includegraphics[width=0.8\textwidth]{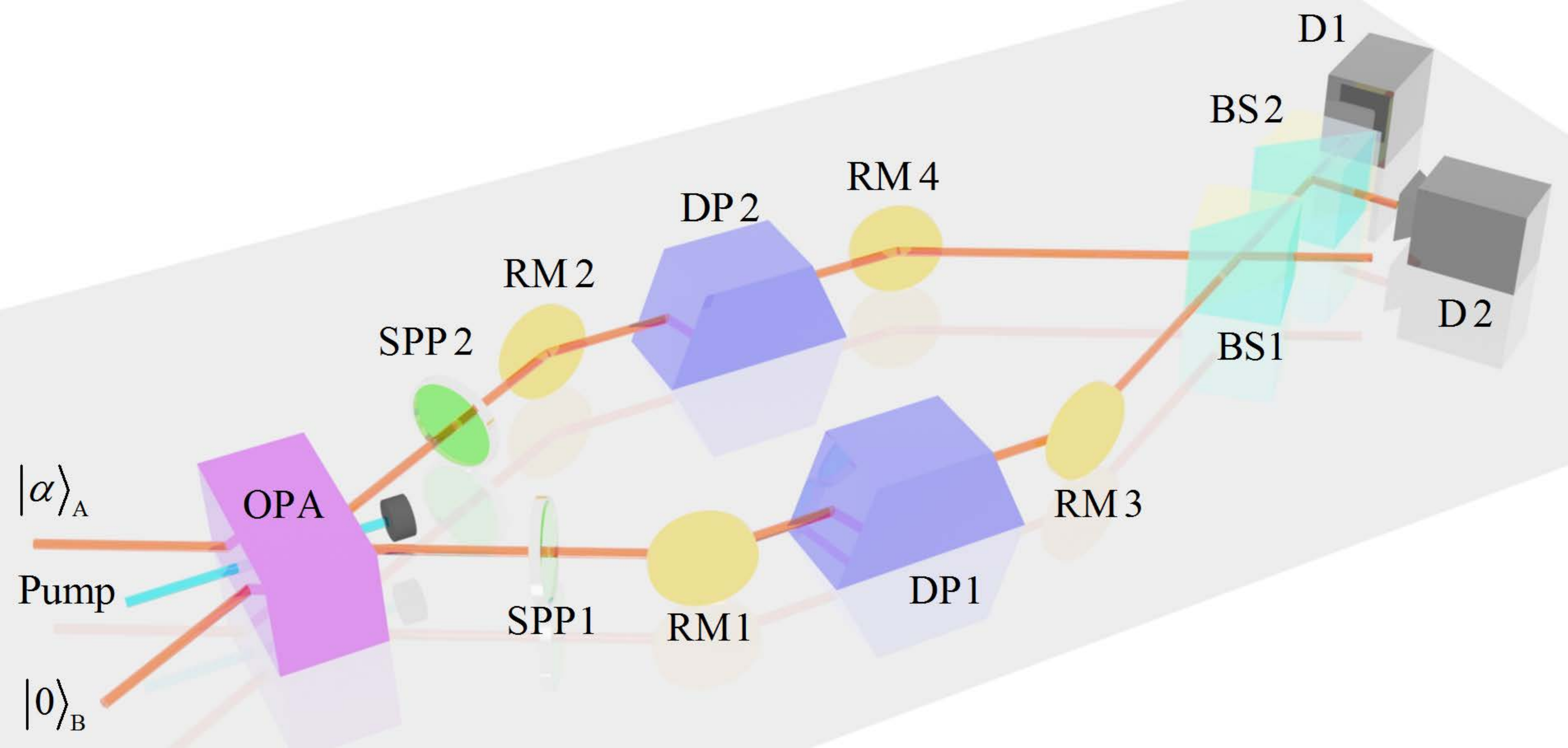}
\caption{A schematic diagram of an OAM-enhanced angular displacement estimation protocol. The input coherent state and pump light enter a hybrid interferometer. Two sets of SPPs and DPs play the roles of modulating OAM and introducing angular displacement respectively. The output signal is detected by two detectors. The devices are abbreviated as: OPA-optical parametric amplifier, RM-reflection mirror, BS-beam splitter, SPP-spiral phase plate, DP-Dove prism, D-detector.}
\label{system}
\end{figure*}

\section{Fundamental principle and device}
\label{II}
Consider a schematic of angular displacement estimation protocol whose the main body is a SU(1,1)-SU(2) hybrid interferometer, as illustrated in Fig. \ref{system}.
This interferometer can be achieved by using an optical parametric amplifier (OPA) to replace the first beam splitter (BS) in SU(2), or, equivalently, utilizing a BS to replace the second OPA in SU(1,1).
The OPA and the BS are used as entry and exit gates of the device respectively. 
A coherent state and a vacuum state are injected into the hybrid interferometer, and beam's OAM degree of freedom is added by two spiral phase plates (SPPs).
The angular displacement difference $\varphi$ between the two Dove prisms (DPs) is the objective we would like to estimate.
It introduces a phase difference $2\ell\varphi$ \cite{1367-2630-13-9-093014} between the two modes, where we assume that OAM quantum number $\ell$ is a positive number.
After the above process, two modes are recombined at BS1, and any output port can be choose to implement a balanced homodyne detection.

Since OPA increases the number of photons, the calculation of the mean photon number $N$ in our protocol is in agreement with the SU (1,1) case \cite{ma2017sub},
\begin{equation}
N = \left\langle {{{\hat A}^\dag }\hat A + {{\hat B}^\dag }\hat B} \right\rangle  = \cosh \left( {2g} \right){\left| \alpha  \right|^2} + 2{\sinh ^2}g.
\end{equation}
Where, 
\begin{eqnarray}
\hat A{\rm{ = }}&&\hat a\cosh g + {\hat b^\dag }\sinh g, \\ 
\hat B{\rm{ = }}&&\hat b\cosh g + {\hat a^\dag }\sinh g,
\end{eqnarray}
the parameter $g$ is the squeezing factor of the OPA, and ${\left| \alpha  \right|^2}$ is the mean photon number of the coherent state before the OPA. 
Here, $ {{\hat a}^\dag }$ ($ {{\hat b}^\dag }$) and $ {\hat a}$ ($ {\hat b}$) are the creation and the annihilation operators for the input modes $A$ ($B$) respectively.

The corresponding SNL and Heisenberg limit (HL) for our protocol are given by
\begin{eqnarray}
 \Delta {\varphi _\textrm{SNL}} &&= \frac{1}{{2\ell\sqrt {\cosh \left( {2g} \right){{\left| \alpha  \right|}^2} + 2{{\sinh }^2}g} }}, \\ 
 \Delta {\varphi _\textrm{HL}} &&= \frac{1}{{2\ell\left[ {\cosh \left( {2g} \right){{\left| \alpha  \right|}^2} + 2{{\sinh }^2}g} \right]}}. 
\end{eqnarray}

One can find that the two limits are boosted by a factor of 2$\ell$. 
This implies that our protocol offers a huge advantage over the protocols without OAM as it is easy to prepare an OAM coherent state with $\ell \le 10$ in the laboratory.
Under this situation, our protocol provides an order of magnitude enhancement over the non-OAM protocols with the same mean photon number. 

From the perspective of metrology, the devices after two DPs can be regarded as measuring devices.
Thus, our protocol can also be considered as a modified measurement protocol based upon SU (1,1) interferometer.
In turn, the ultimate precision defined by quantum Cram\'er-Rao bound (QCRB), or quantum Fisher information (QFI) of our protocol is the same as that of the SU (1,1) interferometer. The QRCB for our protocol is found to be \cite{PhysRevA.93.023810},

\begin{equation}
\Delta {\varphi _\textrm{Q}} = \frac{1}{{2\ell\sqrt {{{\sinh }^2}\left( {2g} \right) + {{\left| \alpha  \right|}^2}\left[ {1 + 2\cosh \left( {2g} \right) + \cosh \left( {4g} \right)} \right]} }}.
\label{Fisher}
\end{equation}

\section{Balanced homodyne detection}
\label{III}
In this section, we analyze balanced homodyne detection, which is a preeminent method for the quantum noise detection by detecting quadrature-phase or quadrature-amplitude.
It can also be used to realize the parity detection towards Gaussian states \cite{plick2010parity}.
Take port $A$ as an example, the detection operators have the following forms
\begin{eqnarray}
 {{\hat X}_{A}} &&= \hat a_\textrm{out} + {{\hat a_\textrm{out}}^\dag }, \\ 
{{\hat P}_{A}} &&= i\left( {{{\hat a_\textrm{out}}^\dag } - \hat a_\textrm{out}} \right).
\end{eqnarray}
Considering $X$ quadrature, we can obtain the expectation value of the $X$ quadrature, 
\begin{equation}
\left\langle {{{\hat X}_{A}}} \right\rangle  =  \sqrt 2 \left| \alpha  \right|\left[ {\cos \left( {\theta  + 2\ell\varphi } \right)\cosh g + \cos \theta \sinh g} \right],
\label{m}
\end{equation}
by using the transformation relationships
\begin{eqnarray}
 {{\hat a}_\textrm{out}} &&= \frac{1}{{\sqrt 2 }}\left[ {\left( {{e^{i2\ell\varphi }}{{\hat a}} + {{\hat b}}} \right)\nu  + \left( {\hat a^\dag  + {e^{i2\ell\varphi }}\hat b^\dag } \right)\mu } \right],
 \label{a1}
 \\ 
 \hat a_\textrm{out}^\dag  &&= \frac{1}{{\sqrt 2 }}\left[ {\left( {{e^{ - i2\ell\varphi }}\hat a^\dag  + \hat b^\dag } \right)\nu  + \left( {{{\hat a}} + {e^{ - i2\ell\varphi }}{{\hat b}}} \right)\mu } \right]. {\kern 1pt}{\kern 1pt}{\kern 1pt}{\kern 1pt}
 \label{a2}
\end{eqnarray}
Where $\nu  = \cosh g$, $\mu  = \sinh g$, and $\theta$ is the amplitude angle of the input coherent state.
Equations (\ref{a1}) and (\ref{a2}) can be derived from the transformation matrices in Appendix \ref{A}. 

\begin{figure}[htbp]
	\centering
	\includegraphics[width=8cm]{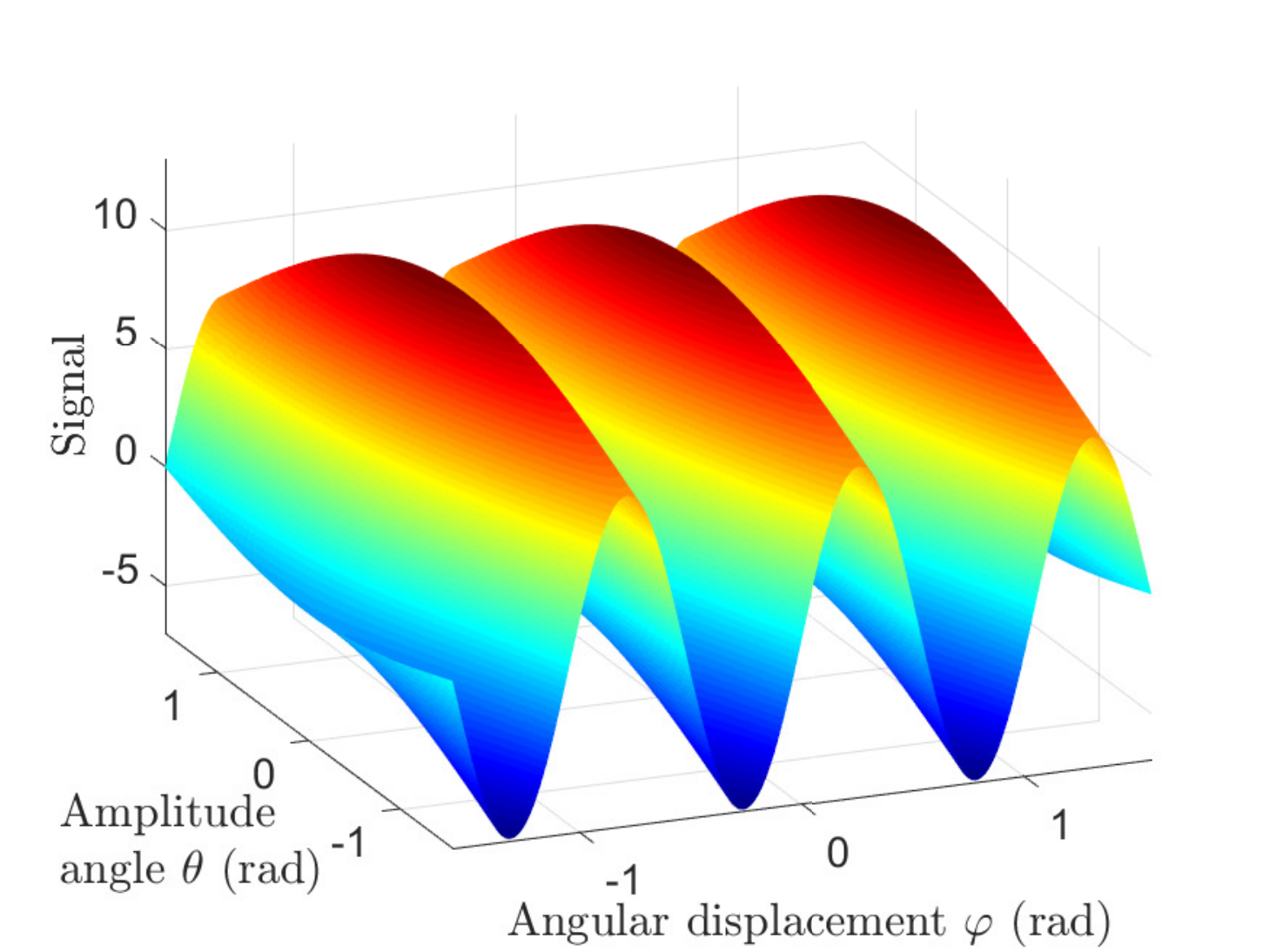}
	\caption{Signal with balanced homodyne detection as a function of angular displacement $\varphi$ (radians) and amplitude angle $\theta$ (radians) in the case of $g=1$, $\ell=3$, $\left|\alpha\right|^2=10$.}
\label{signal}
\end{figure}

We plot the output signal of balanced homodyne detection with $\varphi$ and $\theta$ in Fig. \ref{signal}.
One can see that the signal changes $\ell$ times from $-\pi/2$ to $\pi/2$, i.e., our protocol has 2$\ell$-fold super-resolution characteristic in one period (2$\pi$), like a N00N state.
Furthermore, the signal waveform is also modulated by angle $\theta$.
In terms of the definition of the visibility \cite{doi:10.1080/00107510802091298},
\begin{equation}
V = \frac{{{{\left\langle {{X_{A}}} \right\rangle }_{\max }} - {{\left\langle {{X_{A}}} \right\rangle }_{\min }}}}{{\left| {{{\left\langle {{X_{A}}} \right\rangle }_{\max }}} \right| + \left| {{{\left\langle {{X_{A}}} \right\rangle }_{\min }}} \right|}},
\label{visibility}
\end{equation}
we can find that there is always a 100\% visibility in our protocol, which indicates that $\theta$ has no effect on the visibility of the signal but merely moves the position of the extremums. 

We now turn to another important evaluation criterion, which is the sensitivity in parameter estimation.
Using Eqs. (\ref{a1}) and (\ref{a2}), the expectation value of the square of $X$ quadrature may be written as
\begin{eqnarray}
 \nonumber\left\langle {\hat X_{A}^2} \right\rangle  =&& \cos \left( {2\theta  + 4\ell\varphi } \right){\cosh ^2}g{\left| \alpha  \right|^2} + \cos \left( {2\theta } \right){\sinh ^2}g{\left| \alpha  \right|^2} \\
 \nonumber  &&+ \left[ {\cosh \left( {2g} \right) + \cos \left( {2\ell\varphi } \right)\sinh \left( {2g} \right)} \right]\left( {{{\left| \alpha  \right|}^2} + 1} \right) \\  
  &&+ \cos \left( {2\theta  + 2\ell\varphi } \right)\sinh \left( {2g} \right){\left| \alpha  \right|^2}.
\label{mm}
\end{eqnarray}
In turn, with Eqs. (\ref{m}) and (\ref{mm}), we obtain the fluctuation of $X$ quadrature, as given by
\begin{eqnarray}
\nonumber \Delta {{\hat X}_{A}} =&& \sqrt {\left\langle {\hat X_{A}^2} \right\rangle  - {{\left\langle {{{\hat X}_{A}}} \right\rangle }^2}}  \\ 
  =&& \sqrt {\cosh \left( {2g} \right) + \sinh \left( {2g} \right)\cos \left( {2\ell\varphi } \right)}. 
\end{eqnarray}
Using error propagation, we can calculate the sensitivity of protocol,
\begin{eqnarray}
\nonumber \Delta {\varphi} =&& \frac{{\Delta {{\hat X}_{A}}}}{{\left| {{{\partial \left\langle {{{\hat X}_{A}}} \right\rangle } \mathord{\left/
 {\vphantom {{\partial \left\langle {{{\hat X}_A}} \right\rangle } {\partial \varphi }}} \right.
 \kern-\nulldelimiterspace} {\partial \varphi }}} \right|}} \\ 
=&& \frac{{\sqrt {\cosh \left( {2g} \right) + \sinh \left( {2g} \right)\cos \left( {2\ell\varphi } \right)} }}{{2\sqrt 2 \ell\cosh g\left| {\alpha \sin \left( {\theta  + 2\ell\varphi } \right)} \right|}}. 
\label{bhd}  
\end{eqnarray}

\begin{figure}[htbp]
\centering
\includegraphics[width=8cm]{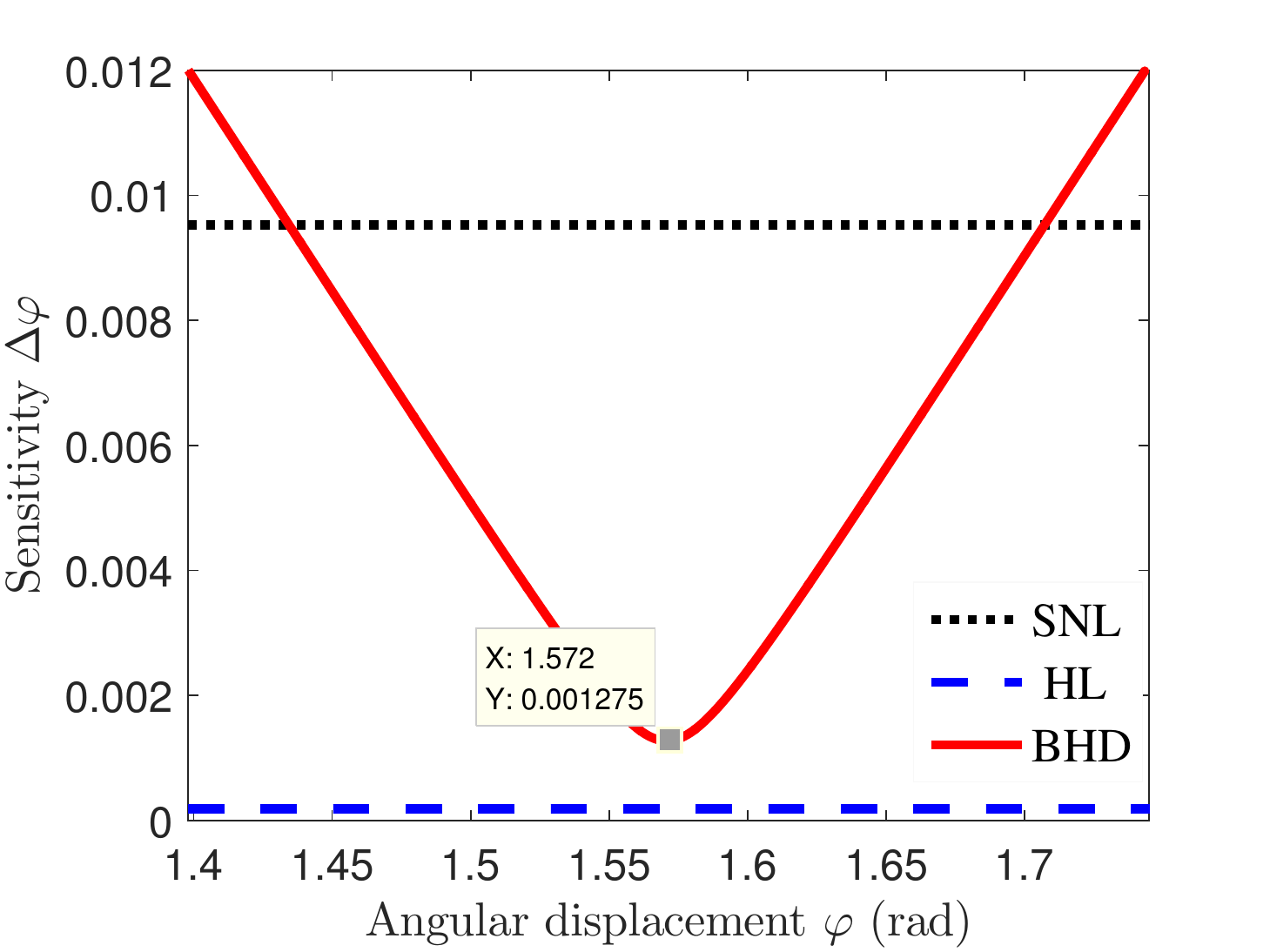}
\caption{Sensitivity with balanced homodyne detection as a function of  angular displacement $\varphi$ (radians) in the case of $g=2$, $\ell=1$ and $\left|\alpha\right|^2=100$.}
\label{bhd_sen}
\end{figure}

To intuitively observe the change in sensitivity, in Fig. \ref{bhd_sen} we plot the sensitivity variation of balanced homodyne detection with angular displacement.
The result manifests that the sensitivity of balanced homodyne detection can surpass SNL and achieve sub-shot-noise-limited sensitivity.
In addition, the variation of the sensitivity curve is slow, this illustrates that quasi-optimal sensitivity can be obtained near the position of optimal sensitivity.
However, it should be explained that the optimal sensitivity in this strategy needs to satisfy the phase matching condition, 
\begin{equation}
{\theta  + 2\ell\varphi }  = k\pi  + {\pi }/{2}
\end{equation}
with an arbitrary integer $k$.

\begin{figure}[htbp]
	\centering
	\includegraphics[width=8cm]{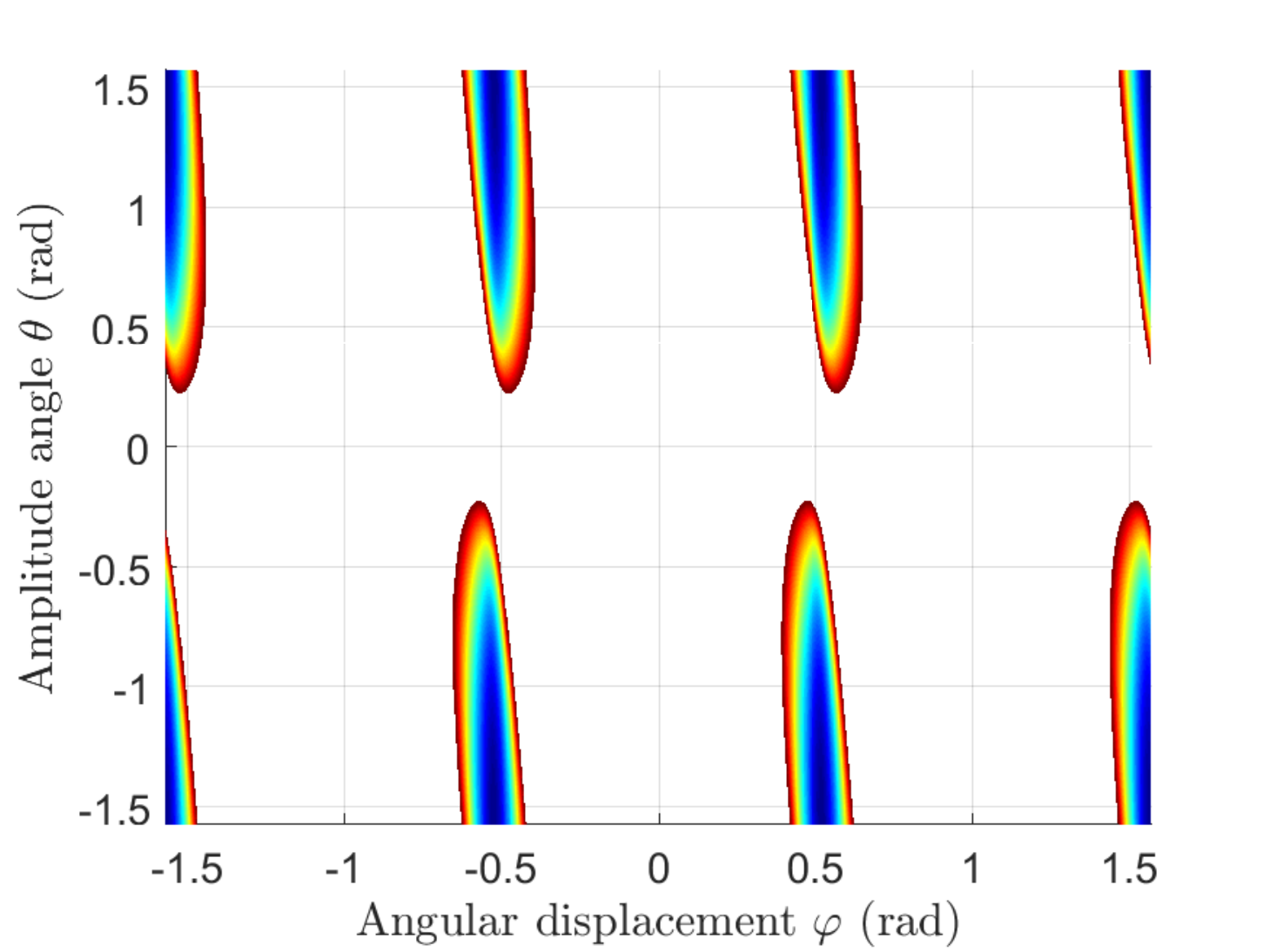}
 	\caption{Sensitivity with balanced homodyne detection as a function of angular displacement $\varphi$ (radians) and amplitude angle $\theta$ (radians) in the case of $g=1$, $\ell=3$, $\left|\alpha\right|^2=10$.}
\label{match}
\end{figure}

In Fig. \ref{match} we plot the impact of $\theta$ on the sensitivity.
For the sake of simplicity, we only plot the regions where the sensitivity is better than SNL (sub-shot-noise-limited regions).
The results reveal that there has 2$\ell$ optimal positions for sensitivity in our protocol.
On the basis of Eq. (\ref{bhd}), we can calculate that optimal sensitivity appears at $2\ell\varphi  = \left( {2k + 1} \right)\pi $ and $2\ell\varphi  + \theta  = k\pi  + {\pi  \mathord{\left/
 {\vphantom {\pi  2}} \right.
 \kern-\nulldelimiterspace} 2}$.
Thus the optimum combination of solution is ${{\left( {2k + 1} \right)\pi } \mathord{\left/
		{\vphantom {{\left( {2k + 1} \right)\pi } {2\ell}}} \right.
		\kern-\nulldelimiterspace} {2\ell}}$ and $\theta  =  \pm {\pi  \mathord{\left/
 {\vphantom {\pi  2}} \right.
 \kern-\nulldelimiterspace} 2}$.

\begin{figure}[htbp]
\centering
\includegraphics[width=8cm]{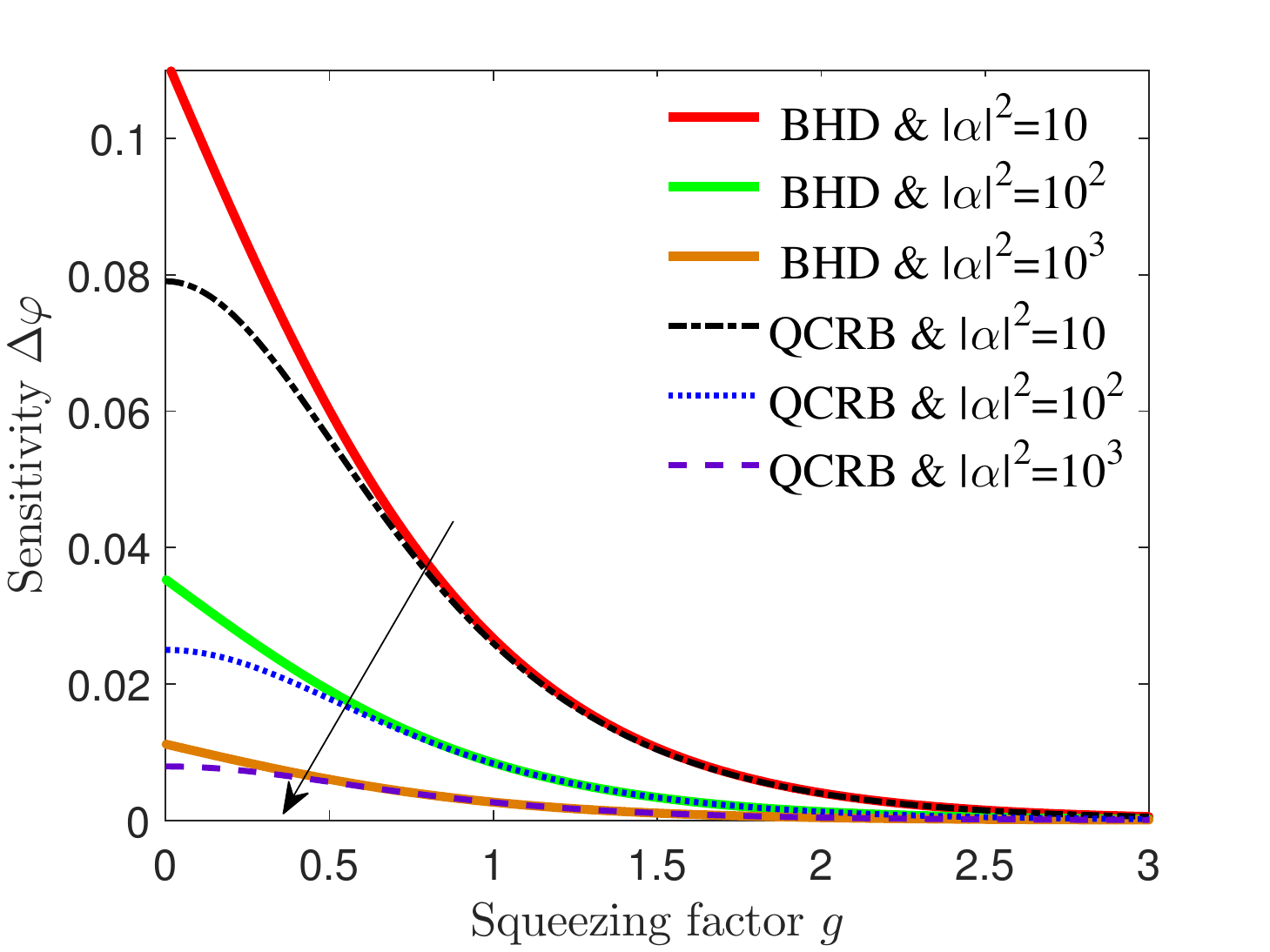}
\caption{Optimal sensitivity with balanced homodyne detection (BHD) as a function of squeezing factor in the case of $\ell=1$ and different values of  $\left|\alpha\right|^2$. The arrow shows the intersection points of sensitivities and QCRBs.}
\label{QCRB}
\end{figure}

For quantificationally analyzing the degree of super-sensitivity in our protocol, we plot Fig. \ref{QCRB} with our sensitivity and QCRB.
We can find that the sensitivity saturates the QCRB with respect to large $g$.
Additionally, with the increase in $\left|\alpha\right|^2$, the trend that sensitivity approaches QCRB quickly can be seen by the pointed arrow in Fig. \ref{QCRB}.
Hence, ${\left| \alpha  \right|^2} \gg 1$ and ${\sinh^2 g} \gg 1$ are two main conditions for saturating the QCRB.
This also implies that balanced homodyne detection is optimal strategy for our protocol.

For the case of ${\left| \alpha  \right|^2} \gg 1$ and ${\sinh^2 g} \gg 1$, we also give a theoretical proof that our protocol always saturates QCRB, i.e., the equivalence between minimum of Eq. (\ref{bhd}) and Eq. (\ref{Fisher}), see Appendix \ref{B} for details.

\section{Analysis and discussion of realistic factor}
\label{IV}

\begin{figure}[htbp]
\centering
\includegraphics[width=8cm]{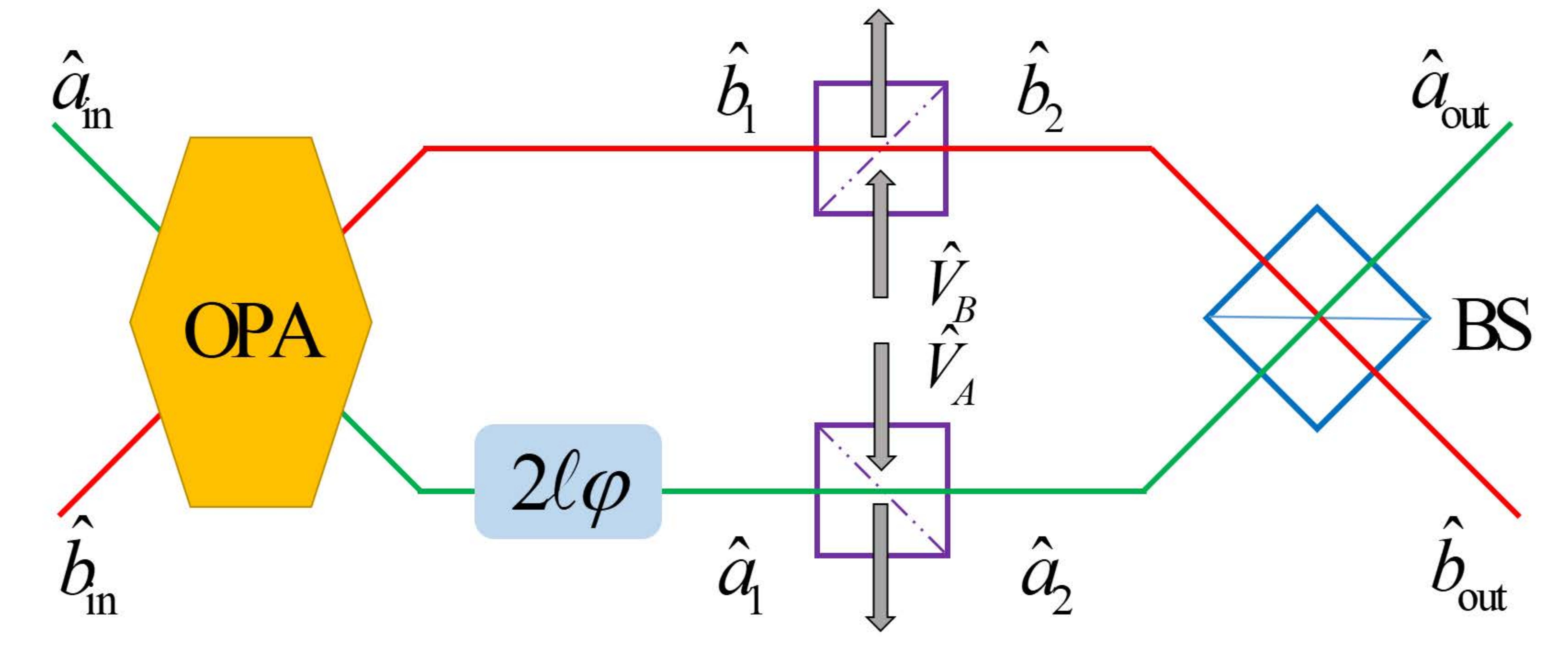}
\caption{A simple model for photon loss. Two virtual BSs are used to simulate photon loss. ${{\hat V}_{A} }$ and ${{\hat V}_{B} }$ are the operators for two vacua.}
\label{floss}
\end{figure}

In this section, we study the effect of a common realistic factor$\---$photon loss$\---$on the sensitivity in our protocol. 
The simple illustration of photon loss is shown in Fig. \ref{floss}.
Suppose that the loss is linear and occurs after the two DPs. 
Such linear loss is usually simulated by placing two virtual beam splitters with arbitrary transmissivity in two arms of the interferometer \cite{1367-2630-16-7-073020}. 
The lost photons are reflected into the environment by the virtual beam splitters.
In the derivation that follows, we assume that the transmissivities of the two virtual beam splitters are $T$.
As the lost photons enter the environment, the number of modes that has to be considered increases from two to four, i.e., two environment ports are added. 
Hence, the transmission matrices change from four-by-four to eight-by-eight, and the details are provided in Appendix \ref{A}. 
According to the transformation relationships, we can obtain the resolution and the sensitivity in the case of photon loss. 

We calculate the expectation values of $X$ quadrature,  
\begin{equation}
{\left\langle {{{\hat X}_{A}}} \right\rangle _\textrm{L}} = \sqrt T \left\langle {{{\hat X}_{A}}} \right\rangle,
\end{equation}
and its square,
\begin{equation}
{\left\langle {\hat X_{A}^2} \right\rangle _\textrm{L}} = T\left\langle {\hat X_{A}^2} \right\rangle  + \left( {1 - T} \right).
\end{equation}
Then, the sensitivity is given by
\begin{equation}
\Delta \varphi _\textrm{L} = \frac{{\sqrt {T\left[ {\cosh \left( {2g} \right) + \sinh \left( {2g} \right)\cos \left( {2\ell\varphi } \right) - 1} \right] + 1} }}{{2\sqrt 2 T\ell\cosh g\left| {\alpha \sin \left( {\theta  + 2\ell\varphi } \right)} \right|}}.
\label{loss2}
\end{equation}

\begin{figure}[htbp]
\centering
\includegraphics[width=8cm]{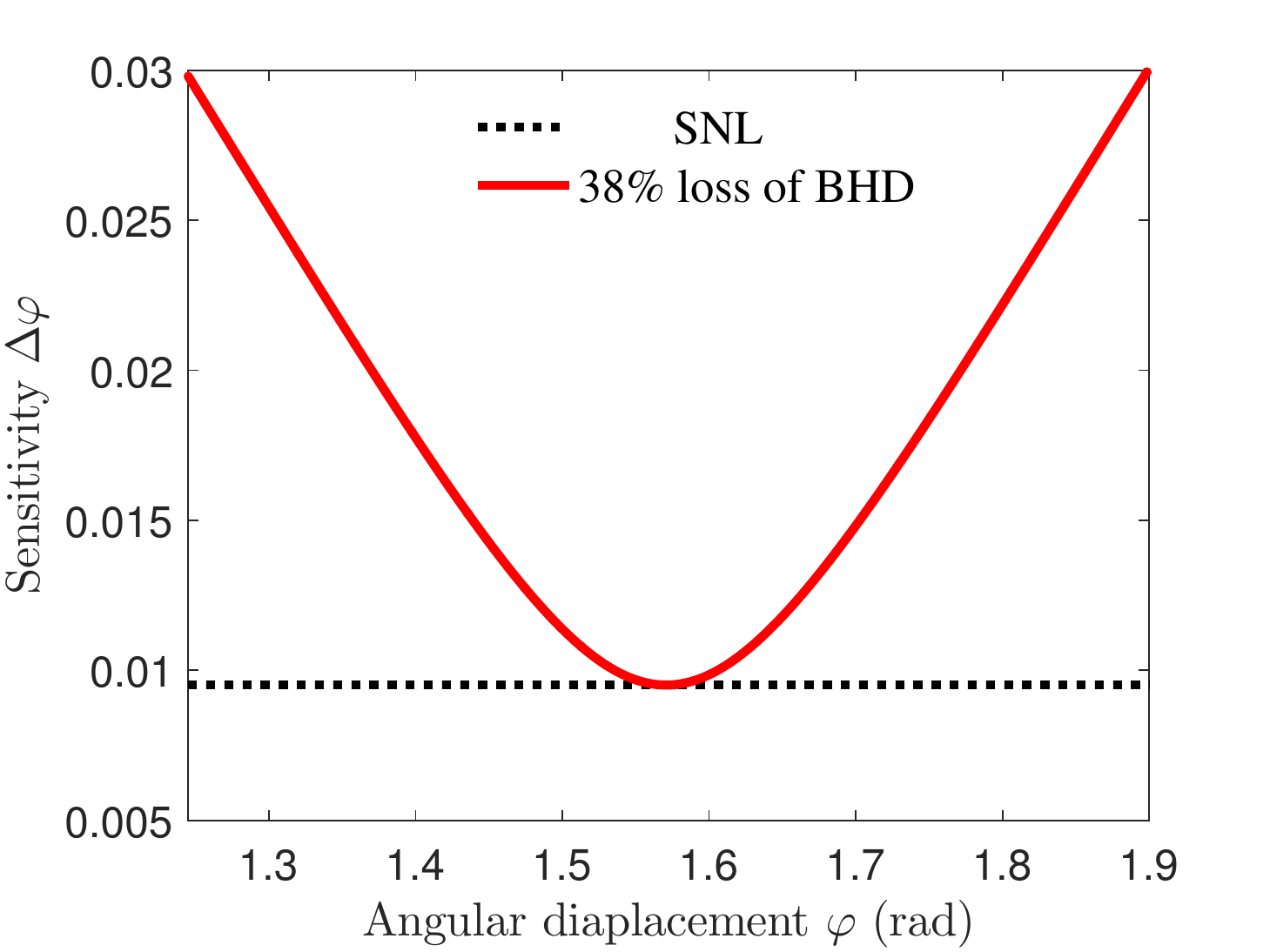}
\caption{Sensitivity with balanced homodyne detection (BHD) as a function of angular displacement $\varphi$ (radians) in the case of $g=2$, $\ell=1$, $\left|\alpha\right|^2=100$ and 38\% loss in transmission process.}
\label{bhd_loss}
\end{figure}

From Eqs. (\ref{visibility}) and (\ref{loss2}), we can see that the visibility is not affected by photon loss, i.e., 100\% visibility can be maintained.
The robustness of balanced homodyne detection is exhibited as there is no change in the signal. 
Figure \ref{bhd_loss} shows the sensitivity of homodyne detection with photon loss.
One observes that our protocol is robust, for it can resist 38\% photon loss in the case of $g=2$, $\ell=1$, and $\left|\alpha\right|^2=100$.
Moreover, a merit is that the change of the sensitivity curve is still slow.

\begin{figure}[htbp]
	\centering
	\includegraphics[width=8cm]{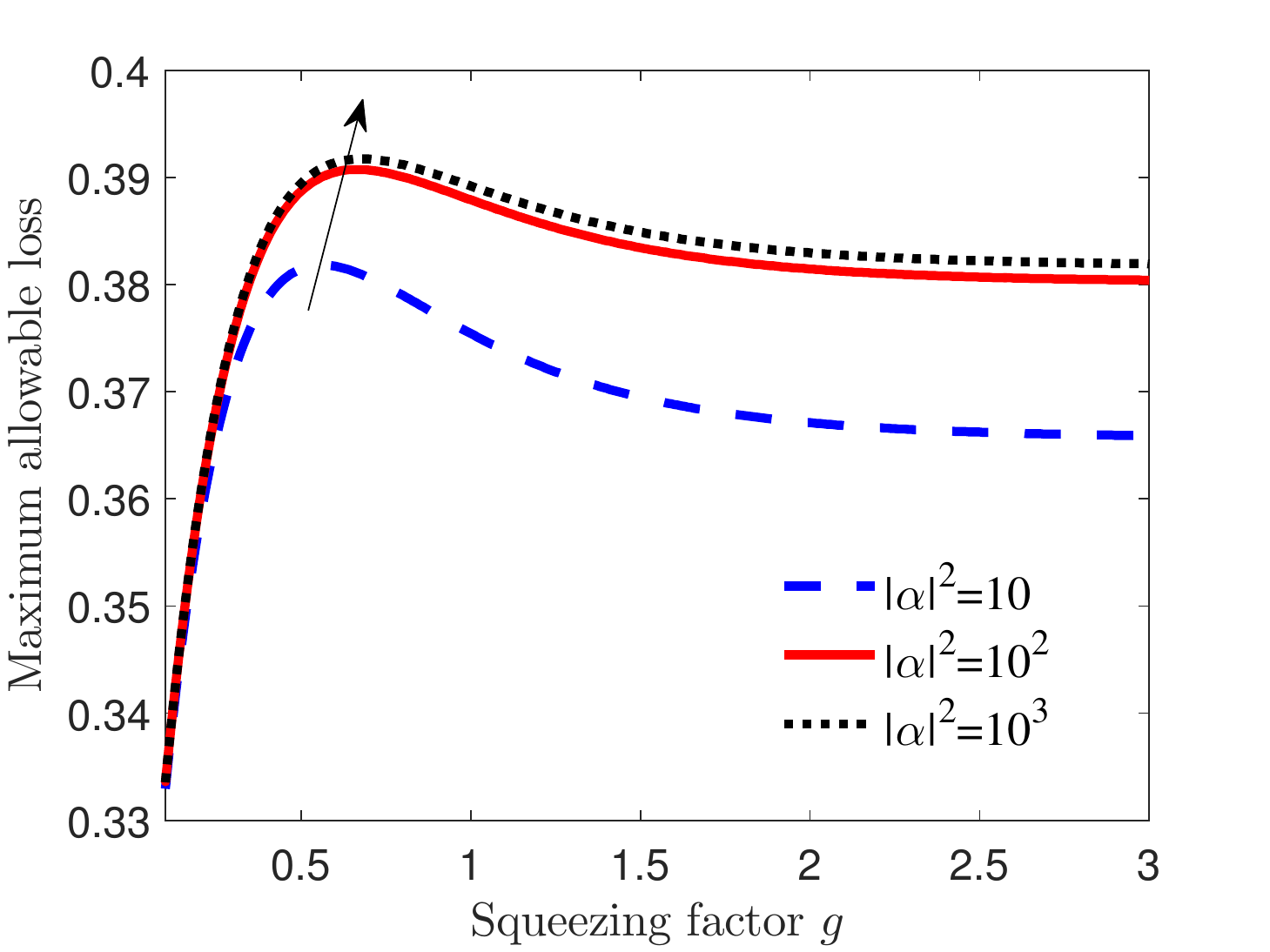}
	\caption{Maximum allowable loss with balanced homodyne detection as a function of squeezing factor in the case of $l=1$ and different mean photon number of coherent state. The arrow shows the maximums of allowable loss curves.}
	\label{maxloss}
\end{figure}

More generally, in Fig. \ref{maxloss}, we discuss the relationship between maximum allowable loss and $g$ with different mean photon number $\left|\alpha\right|^2$.
It can be seen that maximum allowable loss has an optimal position with respect to fixed $\left|\alpha\right|^2$, and the optimal position moves to the right with the increase in $\left|\alpha\right|^2$ (arrow direction).
In addition, with increasing $\left|\alpha\right|^2$, maximum allowable loss also increases and gradually becomes saturated.
An exotic phenomenon is that, no matter what the value of $\left|\alpha\right|^2$ is, the maximum allowable loss ultimately tends to a fixed value when $g$ reaches a threshold value.
Overall, the universal conclusion is that the ability of our protocol possesses an excellent robustness and can resist photon loss (about 30\% to 40\%).

Additionally, we provide a brief discussion for the advantage of our protocol over SU(1,1) interferometer as it is obvious that our protocol is superior than SU(2) \cite{PhysRevA.85.023815}.
For the case of lossless balanced homodyne detection, the optimal sensitivity of Eq. (\ref{bhd}) can be rewritten as 
\begin{equation}
\Delta {\varphi _\textrm{min}} \simeq \frac{1}{{4\ell\cosh g\sqrt {\cosh \left( {2g} \right)} \left| \alpha  \right|}}
\label{theorem}
\end{equation}
with assumptions, ${\left| \alpha  \right|^2} \gg 1$ and ${\sinh^2}g \gg 1$. The details of proof is provided in Appendix \ref{B}.
The sensitivity of a SU(1,1) interferometer with coherent and vacuum states as inputs is given by \cite{PhysRevA.85.023815} : $\Delta {\varphi _{{\rm{SU}}({\rm{1}},{\rm{1}})}} = {1 \mathord{\left/
		{\vphantom {1 {\sqrt {{N_{{\rm{OPA}}}}\left( {{N_{{\rm{OPA}}}} + 2} \right)} \left| \alpha  \right|}}} \right.
		\kern-\nulldelimiterspace} {\sqrt {{N_{{\rm{OPA}}}}\left( {{N_{{\rm{OPA}}}} + 2} \right)} \left| \alpha  \right|}}$.
From Eq. (\ref{theorem}), we can infer that even without OAM's participation, the sensitivity of our protocol is higher than the SU(1,1) by a factor of $\sqrt 2$ under the condition that ${\sinh ^2}g \simeq {\cosh ^2}g=G$ with large $g$ and ${\left| \alpha  \right|^2} \gg 1$.
For coherent state input, our protocol achieves the QCRB that SU(1,1) wants to achieve without success.
Then, the addition of OAM has greatly increased the sensitivity.

Finally, we briefly summarize the merits of OAM. 
Overall, it has played three significant roles in our protocol.
(1) Acting as the linear amplifier for angular displacement, mathematically, equivalent to the increase of the repeated trials in an actual measurement.
(2) Extending the super-resolution output signal from a single to 2$\ell$-fold.
(3) The multiple positions of optimal sensitivity provided with OAM make the scanning range in the actual detection shortened to ${1 \mathord{\left/
		{\vphantom {1 {2l}}} \right.
		\kern-\nulldelimiterspace} {2\ell}}$.

\section{Conclusion}
\label{V}
In summary, we present a protocol for estimating angular displacement using OAM coherent state and a hybrid interferometer. 
Balanced homodyne detection is studied, and the results show that both super-resolution and super-sensitivity are achieved in lossless scenario. 
The output signal has a 100\% visibility, and we demonstrate that a sub-shot-noise-limited sensitivity, which is saturated by QCRB can be obtained when both $\left| \alpha\right|^2 $ and $\sinh^2 g$ are much larger than 1.
Additionally, we explore the effects of photon loss on the resolution and sensitivity.
The 100\% visibility is maintained in loss scenario, and our protocol can resist photon loss of more than 30\%.
We briefly discuss the advantage of our protocol compared to the SU(1,1) and show that the sensitivity of our protocol improves with a factor of $\sqrt 2$. 
Using OAM, the enhanced effect is reflected by a factor of $2\sqrt 2\ell$ boost in sensitivity compared with non-OAM protocols.
Finally, the merits of taking advantage of OAM is summarized.

\section*{Acknowledgments} 
This work is supported by the National Natural Science Foundation of China (Grant No. 61701139).

\section*{appendix} 
\appendix

\section{Transformation matrices of optical processes for the phase space}
\label{A}

In this part of the Appendix, we provide the transformation matrices for the optical processes in the phase space.
The matrices for OPA, angular displacement and BS are given by 
\begin{equation}
{\mathbf{U}_\textrm{OPA}}= \left( {\begin{array}{*{20}{c}}
   {\cosh g} & 0 & {\sinh g} & 0  \\
   0 & {\cosh g} & 0 & { - \sinh g}  \\
   {\sinh g} & 0 & {\cosh g} & 0  \\
   0 & { - \sinh g} & 0 & {\cosh g}  \\
\end{array}} \right),
\end{equation}
\begin{equation}
{\mathbf{U}_\textrm{AD}} = \left( {\begin{array}{*{20}{c}}
   {\cos \left( {2\ell\varphi } \right)} & { - \sin \left( {2\ell\varphi } \right)} & 0 & 0  \\
   {\sin \left( {2\ell\varphi } \right)} & {\cos \left( {2\ell\varphi } \right)} & 0 & 0  \\
   0 & 0 & 1 & 0  \\
   0 & 0 & 0 & 1  \\
\end{array}} \right),
\end{equation}
\begin{equation}
{\mathbf{U}_\textrm{BS}} = \frac{1}{{\sqrt 2 }}\left( {\begin{array}{*{20}{c}}
   1 & 0 & 1 & 0  \\
   0 & 1 & 0 & 1  \\
   { - 1} & 0 & 1 & 0  \\
   0 & { - 1} & 0 & 1  \\
\end{array}} \right).
\end{equation}

For the case of photon loss, the transformation matrices transforms from four-by-four to eight-by-eight, a rise in dimensionality due to the introduction of the environment modes. The specific forms are as follows:
\begin{equation}
{\mathbf{\tilde{U}}_\textrm{OPA}} = {\mathbf{U}_\textrm{OPA}} \oplus {\mathbf{I}\left( 4 \right)},
\end{equation}
\begin{equation}
{\mathbf{\tilde{U}}_\textrm{AD}} = {\mathbf{U}_\textrm{AD}} \oplus {\mathbf{I}\left( 4 \right)},
\end{equation}

\begin{equation}
{\mathbf{\tilde{U}}_\textrm{VBS}} = {\left( {\begin{array}{*{20}{c}}
   {\sqrt T  {\kern 1pt}{\kern 1pt}  \mathbf{I}\left( 4 \right)} & {\sqrt {1 - T}  {\kern 1pt}{\kern 1pt} \mathbf{I}\left( 4 \right)}  \\
   {\sqrt {1 - T}  {\kern 1pt}{\kern 1pt} \mathbf{I}\left( 4 \right)} & { - \sqrt T  {\kern 1pt}{\kern 1pt} \mathbf{I}\left( 4 \right)}  \\
\end{array}} \right)_{8 \times 8}}
\end{equation}

\begin{equation}
{\mathbf{\tilde{U}}_\textrm{BS}} ={\mathbf{U}_\textrm{BS}} \oplus {\mathbf{I}\left( 4 \right)}.
\end{equation}
Where ${\mathbf{I}\left( 4 \right)}$ is a four-by-four identity matrix, and $\oplus$ is direct sum.

\section{The proof of equivalence between optimal sensitivity of Eq. (\ref{bhd}) and QCRB (Eq. (\ref{Fisher})), and the proof process of Eq. (\ref{theorem})}
\label{B}

In this section, we start off the equivalence between optimal sensitivity (Eq. (\ref{bhd})) and QCRB (Eq. (\ref{Fisher})), i.e., we only need to give the derivation of ${{\Delta {\varphi _{\min }}} \mathord{\left/
{\vphantom {{\Delta {\varphi _{\min }}} {\Delta {\varphi _\textrm{Q}}}}} \right.	\kern-\nulldelimiterspace} {\Delta {\varphi _Q}}} = 1$.
By virtue of ${\left| \alpha  \right|^2} \gg 1$, we have
\begin{eqnarray}
\nonumber\frac{{\Delta {\varphi _{\min }}}}{{\Delta {\varphi _Q}}} &&= {{\frac{{\sqrt {\cosh \left( {2g} \right) - \sinh \left( {2g} \right)} }}{{\sqrt 2 \cosh g}}}}{{{\sqrt {1 + 2\cosh \left( {2g} \right){\rm{ + }}\cosh \left( {4g} \right)} }}} \\ 
\nonumber&&= \frac{{\sqrt {\cosh \left( {2g} \right) - \sinh \left( {2g} \right)} }}{{\cosh g}}\sqrt {\cosh \left( {2g} \right)\left[ {1 + \cosh \left( {2g} \right)} \right]}  \\ 
\nonumber&&\simeq \frac{{\sqrt {1 + \cosh \left( {2g} \right)} }}{{\sqrt 2 \cosh g}} \\ 
&&= 1. 
\end{eqnarray}
with $2\cosh \left( {2g} \right) \simeq {e^{2g}} \simeq 2\sinh \left( {2g} \right)$ and ${e^{ - 2g}} \simeq 0$ for large $g$.

Next, we consider Eq. (\ref{theorem}) with the same approximation, the proof is given by
\begin{widetext}
\begin{equation}
\Delta {\varphi _{\min }} = \frac{{\sqrt {\cosh \left( {2g} \right) - \sinh \left( {2g} \right)} }}{{2\sqrt 2 \ell\cosh g\left| \alpha  \right|}} = \frac{1}{{2\ell\sqrt {2{e^{2g}}} \cosh g\left| \alpha  \right|}} \simeq \frac{1}{{4\ell\cosh g\sqrt {\cosh \left( {2g} \right)} \left| \alpha  \right|}}.
\end{equation}
\end{widetext}

%\bibliography{bibfile}

%

\end{document}